\documentclass{aastex631}
\usepackage{CJK}

\usepackage{graphicx}	
\usepackage{amsmath}	
\usepackage{amssymb}	
\usepackage{float}
\usepackage{enumerate}
\usepackage{color}
\usepackage{subfigure}
\usepackage{graphicx}
\usepackage{booktabs}
\usepackage{threeparttable}

\usepackage{enumitem}

\submitjournal{ApJS}

\shorttitle{Be star identification}
\shortauthors{Tan et al.}

\graphicspath{{./}{figures/}}

\begin{document}
\begin{CJK*}{UTF8}{gbsn}

\begin{sloppypar}

\title{A robust method for identifying Be stars in the LAMOST Data Release 11 \\ based on Deep-learning approach}
\author[0000-0001-6215-9242]{Lei Tan({\CJKfamily{gbsn}谈磊})}
\affiliation{Center for Astrophysics and Great Bay Center of National Astronomical Data Center, \\ Guangzhou University, Guangzhou, China, 510006}

\author[0000-0002-8765-3906]{Hui Deng ({\CJKfamily{gbsn}邓辉})}
\affiliation{Center for Astrophysics and Great Bay Center of National Astronomical Data Center, \\ Guangzhou University, Guangzhou, China, 510006}

\author[0000-0002-7960-9251]{Ying Mei({\CJKfamily{gbsn}梅盈})}
\affiliation{Center for Astrophysics and Great Bay Center of National Astronomical Data Center, \\ Guangzhou University, Guangzhou, China, 510006}


\author[0000-0001-7343-7332]{Huanbin Chi({\CJKfamily{gbsn}迟焕斌})}
\affiliation{School of Media and  Information Engineering, \\ Yunnan Open University, Kunming, China, 650599}



\author[0009-0004-9075-4208]{Yixing Chen({\CJKfamily{gbsn}陈艺星})}
\affiliation{Center for Astrophysics and Great Bay Center of National Astronomical Data Center, \\ Guangzhou University, Guangzhou, China, 510006}

\author[0009-0004-7727-4613]{Tianhang Liu({\CJKfamily{gbsn}刘天航})}
\affiliation{Center for Astrophysics and Great Bay Center of National Astronomical Data Center, \\ Guangzhou University, Guangzhou, China, 510006}

\author[0000-0002-9847-7805]{Feng Wang({\CJKfamily{gbsn}王锋})}
\affiliation{Center for Astrophysics and Great Bay Center of National Astronomical Data Center, \\ Guangzhou University, Guangzhou, China, 510006}


\correspondingauthor{ Hui Deng, Ying Mei, Feng Wang}
\email{denghui@gzhu.edu.cn,meiying@gzhu.edu.cn,fengwang@gzhu.edu.cn}

\begin{abstract}
Be stars are rapidly rotating B-type stars that exhibit Balmer emission lines in their optical spectra. These stars play an important role in studies of stellar evolution and disk structures. 
In this work, we carried out a systematic search for Be stars based on LAMOST spectroscopic data. Using low-resolution spectra from LAMOST DR11, we constructed a data set and developed a classification model that combines long short-term memory networks and convolutional neural networks , achieving a testing accuracy of 97.86\%. 
The trained model was then applied to spectra with signal-to-noise ratios greater than 10, yielding 55,667 B-type candidates. With the aid of the MKCLASS automated classification tool and manual verification, we finally confirmed 40,223 B-type spectra. 
By cross-matching with published H$\alpha$ emission-line star catalogs, we obtained a sample of 8298 Be stars, including 3787 previously reported Be stars and 4511 newly discovered. 
Furthermore, by incorporating color information, we classified the Be star sample into Herbig Be stars and Classical Be stars.  In total, we identified 3363 Classical Be stars and 35 Herbig Be stars.
The B-type and Be star catalogs derived in this study, together with the code used for model training, have been publicly released to facilitate community research.

\end{abstract}

\keywords{Emission line stars(460); Astronomy data analysis(1858); Stellar spectral types(2051); Stellar classification(1589); Convolutional neural networks(1938);
}

\section{Introduction}
Be stars are a class of B-type stars that exhibit Balmer emission features in their optical spectra, usually associated with a gaseous circumstellar disk \citep{Porter2003}. According to their formation mechanisms and evolutionary stages, Be stars are generally divided into two categories: Herbig Be stars (HBes), which are typically in the early stages of stellar formation and whose emission features originate from an accretion disk surrounding the star \citep{Herbig1960,Waters1998}; and Classical Be stars (CBes), which are rapidly rotating B-type stars in the main-sequence or giant phases, with emission lines primarily arising from a rotationally supported disk formed by stellar mass ejection \citep{Rivinius2013}. The distinction between these two types often relies on differences in color–magnitude information, infrared excess, and spectral features \citep{Zorec1997,Vieira2003}. The infrared excess of HBes is much larger than that of CBes \citep{Finkenzeller1984}. As important objects for investigating rapid stellar rotation and disk structures, Be stars provide key samples for studying stellar formation and evolution, and serve as crucial laboratories for understanding stellar angular momentum loss as well as the formation and dissipation mechanisms of circumstellar disks \citep{Porter2003,Rivinius2013}.


A systematic search for Be stars is of great significance for understanding their roles in stellar evolution, angular momentum transfer, and disk formation. In previous studies, researchers mainly relied on color–magnitude diagrams, H$\alpha$ narrow-band photometric measurements, and spectroscopic identification to search for Be stars. For example,\citet{Sabogal2008} searched for Be star candidates in the OGLE II Galactic bulge fields using color-magnitude diagrams combined with variability information, and identified 29,053 candidates, among which about 1,500 were classified as reliable Be stars. 
\citet{Vieira2003} present a catalog of 108 Herbig Ae/Be candidate stars identified in the Pico dos Dias Survey, together with 19 previously known candidates and four objects selected from the IRAS Faint Source Catalog through combined optical and infrared color analysis. 
\citet{McSwain2005} employed H$\alpha$ narrow-band photometry to search for Be stars in 55 southern open clusters, and identified 52 definite Be stars along with an additional 129 candidates.
\citet{Mathew2008} conducted a slitless spectroscopic survey of 207 Galactic open clusters and identified 157 emission-line stars, of which 98\% were classified as Be stars.
\citet{Chojnowski2015} employed APOGEE high-resolution near-infrared spectroscopy and identified 128 newly found CBes based on Brackett emission lines. 
Using IPHAS photometry and an automated color–color selection method, \citet{Witham2008} constructed a catalogue of 4853 Hα emission-line sources in the northern Galactic plane, of which about 70\% were later confirmed to be CBes by \citet{Gkouvelis2016}.
However, traditional approaches have limitations. First, methods based on color or narrow-band photometry are easily affected by extinction, infrared excess, and photometric uncertainties, leading to misidentifications or omissions.
Second, spectroscopic identification relying on manual inspection is inefficient and cannot cope with the millions of spectra produced by current large-scale surveys.
Therefore, in the context of large-scale survey data, it is imperative to develop efficient, automated, and robust methods for Be star identification.

Large-scale spectroscopic surveys provide abundant data resources for for searching for Be stars. Leveraging their advantages in automatic feature extraction and pattern recognition, machine learning and deep learning techniques are increasingly becoming powerful tools for stellar classification and the identification of peculiar objects.
In the identification of Be stars or emission-line stars, several applications based on large survey samples have been reported. 
\citet{Mohr2017} employed VPHAS+ photometry to detect 14,900 massive OB stars in the Carina Arm. Subsequently, using a neural network classifier, \citet{Aidelman2020} confirmed 248 of these as CBe stars. 
\citet{Perez2017} used machine-learning classifiers on OGLE I-band light curves in the Large Magellanic Cloud and identified 50 CBe candidates.  
\citet{Vioque2020} applied a supervised machine-learning model to Gaia DR2 photometry to search for Herbig Ae/Be stars, and reported 693 new CBe candidates across the Galactic plane.  
\citet{Cotar2021} applied a neural network autoencoder to GALAH survey data and identified 10,364 stellar spectra exhibiting emission components.

The Large Sky Area Multi-Object Fiber Spectroscopic Telescope (LAMOST; \citealt{cui2012large, zhao2012lamost, deng2012lamost}) is a reflecting Schmidt telescope with a large aperture and wide field of view, capable of placing 4,000 fibers on its focal plane to simultaneously observe 4,000 targets. Its low-resolution spectroscopic survey covers 3700--9000\,\AA\ with a resolution of $R \sim 1800$ at 5500\,\AA, and the data release 11 (DR11) provides 11,944,049 calibrated spectra, including 11,586,067 stellar spectra. In addition, LAMOST also conducts a medium-resolution survey ($R \sim 7500$) in the blue (4950--5350\,\AA) and red (6300--6800\,\AA) bands, which has yielded  36,933,882 single exposure spectra and 10,005,645 coadded spectra. These large spectroscopic data sets offer unprecedented opportunities for systematic studies of stellar populations and for searching special types of stars such as Be stars and emission-line stars.
For example, 
\citet{Lin2015} applied automated classification to LAMOST DR1 low-resolution spectra and detected 192 CBe candidates.  
\citet{Hou2016} constructed a catalog of 11,204 emission-line spectra from LAMOST DR2, including 5594 CBe candidates and 23 HBe candidates.  
\citet{Skoda2020} employed an active-learning classification approach on LAMOST DR2 and identified 948 new emission-line stars corresponding to 1013 spectra.  
\citet{Anusha2021} analyzed data from LAMOST DR5 and reported 159 classical Ae stars.  
\citet{Wang2022} used the LAMOST DR7 MRS database and an 18-layer ResNet deep convolutional neural network to identify 1,162 Be candidates, ultimately yielding 183 new CBes.  
\citet{Zhang2022} analyzed 30,023 spectra of 25,867 early-type emission-line stars from LAMOST DR7, classified their H$\alpha$ profile morphologies, and applied updated color-color criteria, yielding 2,636 CBe candidates and 118 HBe candidates, including 62 newly identified.

In light of the previous literature, we designed a framework that integrates bidirectional long short-term memory networks (BiLSTM) with convolutional neural networks (CNN), and trained a model to identify B-type stars. The Be stars were then recognized by cross-matching the identified B-type stars with H$\alpha$ emission-line star catalogs. Finally, the Be star candidates were visually inspected and classified into CBe and HBe stars according to specific criteria.  
Section \ref{2} describes the data used in this study and the preprocessing methods. Section \ref{3} introduces the model architecture, training process, and the procedures for obtaining B-type and Be stars. Section \ref{4} presents the classification and verification of Be stars, while Section \ref{5} analyzes the model performance and the properties of our B-type and Be star samples. Finally, Section \ref{6} summarizes the main findings, discusses the current limitations, and outlines future research directions.

\section{Data and Preprocessing}\label{2}
The spectroscopic data used in this work are drawn from the LAMOST DR11 low-resolution survey. To ensure data quality, we adopted the signal-to-noise ratio (S/N) provided by the LAMOST pipeline and selected spectra with $g$-band S/N $\geq 10$ as our data source.  

To construct the training and testing data for B-type stars, we cross-matched DR11 with the B-type stars published by \citet{Liu2019}, yielding 21,056 B-type spectra (including a small fraction of O-type spectra). Among these, 5,000 spectra were randomly selected as the B-type samples of our data set. In addition, to maintain discriminative ability across multiple spectral types and to mitigate class imbalance, we randomly selected 5,000 spectra from each of the other five categories (A, F, G, K, and M). The M-type sample includes both M-type dwarfs and M-type giants.  

\begin{figure}[htbp]
    \centering
    \includegraphics[width=0.8\linewidth]{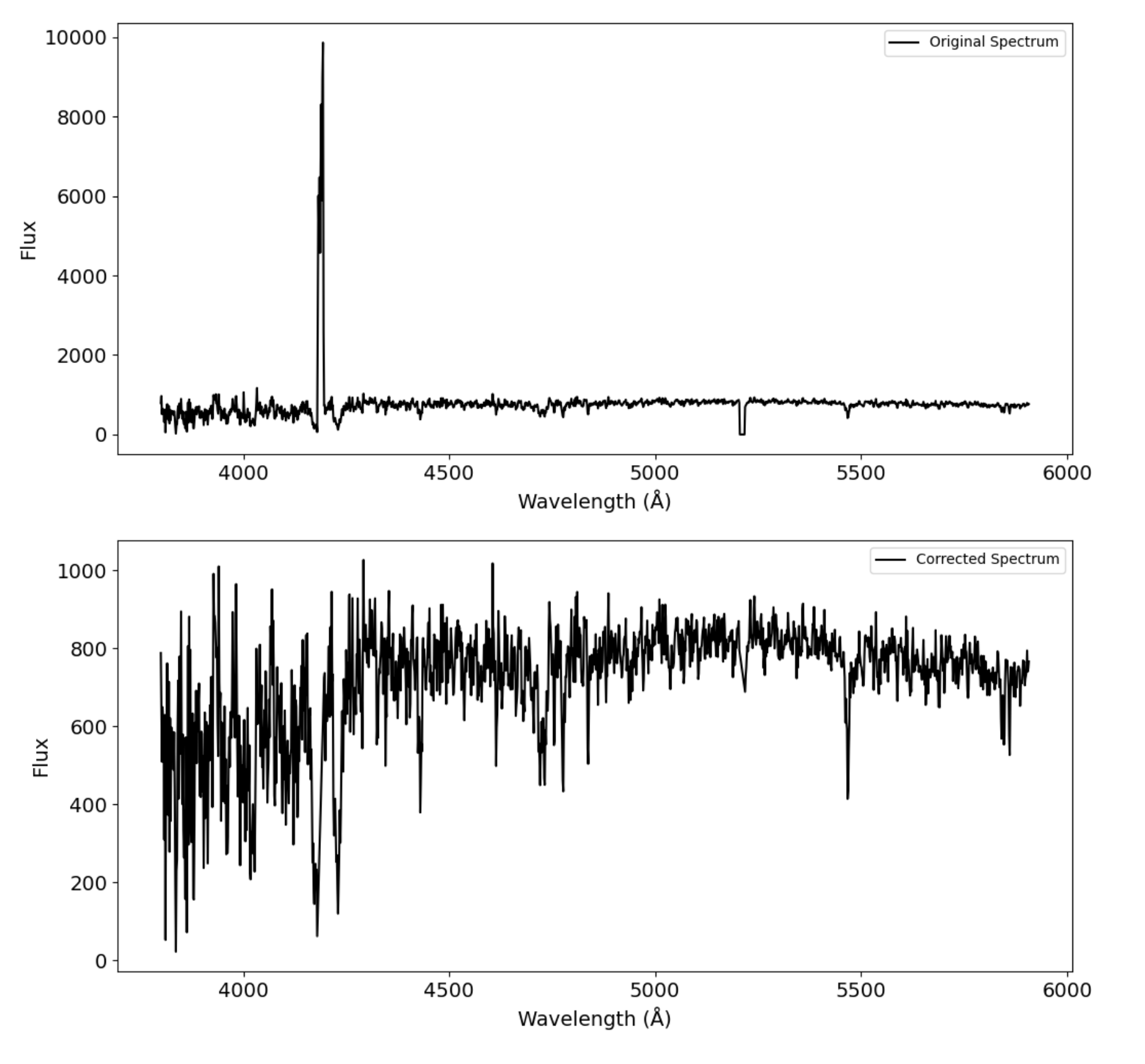}
    \caption{An example of cosmic-ray removal and missing-value correction. The green line represents the original spectrum, while the orange line shows the spectrum after processing.}
    \label{fig:cosmos}
\end{figure}

Spectra with missing values or cosmic-ray contamination can interfere with model training and prediction, potentially leading to misclassification. Therefore, we performed preprocessing on these spectra. For spectra with missing values, if the number of consecutive missing pixels was $\leq 10$, we filled them using local linear interpolation; if the number exceeded 10, the spectrum was discarded. For spectra affected by cosmics, we first performed continuum fitting and normalization. Abnormally high-flux pixels were then identified by comparing the normalized flux values with a threshold set to 1.5 (in normalized flux units), and these pixels were regarded as contaminated by cosmic-rays. The affected pixels were subsequently replaced with interpolated values from adjacent valid pixels to correct the flux. The effects of missing-value treatment and cosmic-ray correction are shown in Figure \ref{fig:cosmos}.

\begin{figure*}[htbp]
\begin{center}
\includegraphics[width=0.8\linewidth]{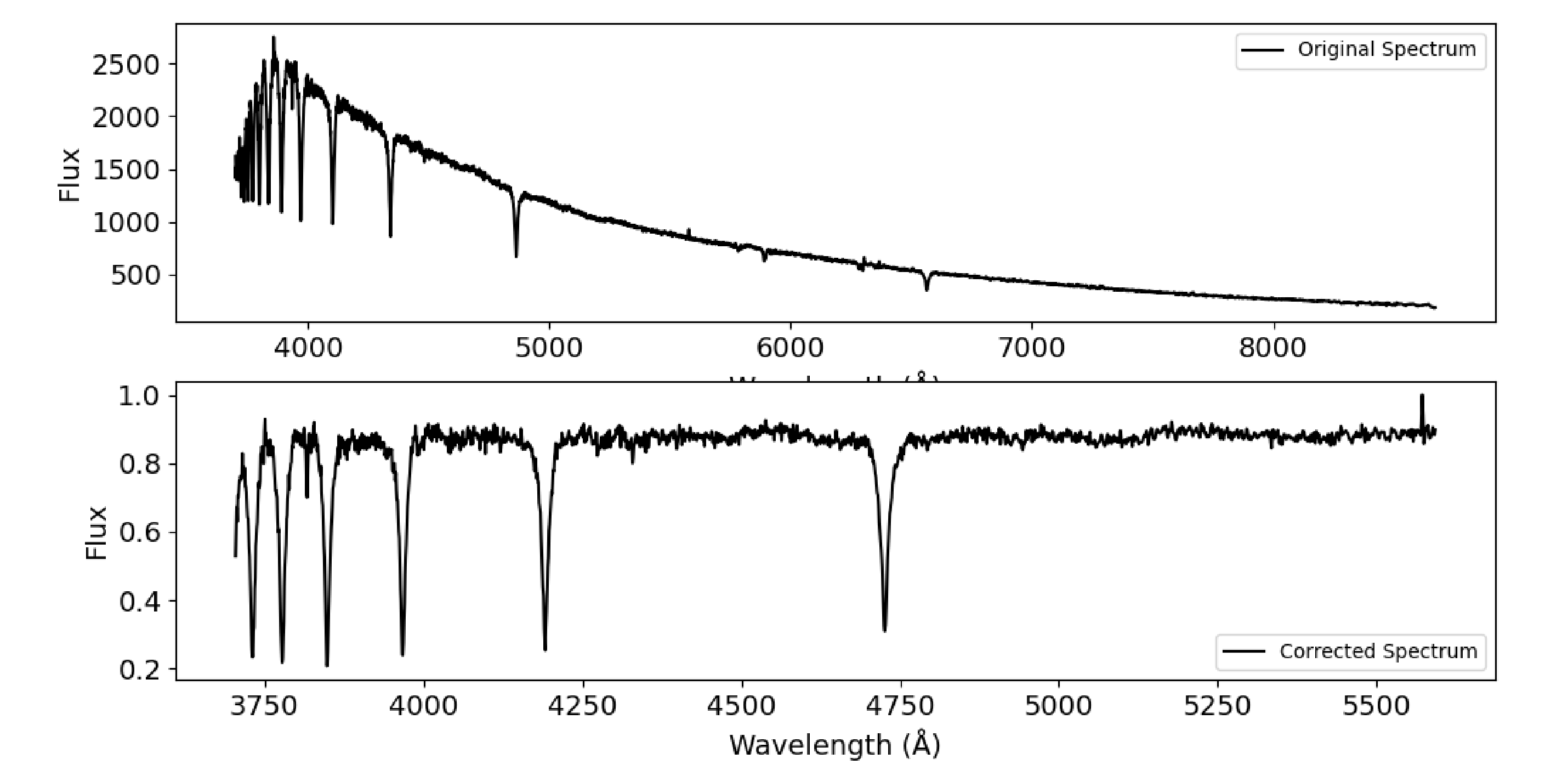}
\end{center}
\caption{Comparison of spectra before and after processing. The upper panel shows the original spectrum, while the lower panel presents the spectrum after truncation and cubic spline continuum fitting and normalization.}
\label{cubic}
\end{figure*}

The LAMOST low-resolution spectra cover the wavelength range $\sim 3700$--$9000$\,\AA. Considering that early-type stars exhibit abundant Balmer and He\,\textsc{i} absorption/emission features in the blue region, we only used the first 1,800 flux values of each spectrum (corresponding to $\sim 3700$--$5600$\,\AA) in the preprocessing stage as input. This range encompasses H$\delta$ (4102\,\AA), H$\gamma$ (4340\,\AA), H$\beta$ (4861\,\AA), and multiple He\,\textsc{i} lines, which are crucial for distinguishing early-type stars (O and B) from later types (A, F, G, K, M). At the same time, this reduces interference from unrelated features. Finally, we applied cubic spline continuum fitting and normalization to the flux within the truncated wavelength range, and then all samples were normalized using Equation \ref{eq:normal}. The processed result is illustrated in Figure \ref{cubic}.

\begin{equation}
    {x}'=\frac{x - x_{min}}{x_{max} - x_{min}}
	\label{eq:normal}
\end{equation}

\section{Method}\label{3}
This section describes the procedure used for Be star identification. Considering the large diversity in the H$\alpha$ emission-line profiles among Be stars (in both shape and intensity), identifying Be stars directly from full spectra is challenging, as those with weak H$\alpha$ emission line are easily overlooked. To address this, we first trained a B-type star classification model to identify B-type stars from the low-resolution spectra. The B-type stars identified by the model were then cross-matched with published catalogs of H$\alpha$ emission-line stars to obtain Be star candidates. 

\subsection{B-type star identification model}\label{3.1}
Our data set consists of 30,000 spectra, with 5,000 samples in each category. After preprocessing, the data set was randomly divided into training and testing sets, with 70\% used for training and 30\% for testing. The training set was employed to optimize model parameters, while the testing set was used to evaluate model performance on independent samples, ensuring the generalization ability of the model.  

In this study, we adopted a hybrid architecture that combines bidirectional long short-term memory networks (BiLSTM) and convolutional neural networks (CNN). 
LSTM\citep{Hochreiter1997}, as a recurrent neural network capable of capturing long-range dependencies in sequential data, is particularly suitable for handling one-dimensional spectral sequences. The bidirectional structure (BiLSTM), integrating the advantages of BiRNN \citep{Schuster} and LSTM, enables simultaneous learning of forward and backward contextual information from spectra, leading to more comprehensive sequence modeling. CNNs excel at extracting local features and are effective in identifying spectral line features (e.g., H$\alpha$, H$\beta$, He\,\textsc{i}). Through convolution and max-pooling operations, CNNs are robust to noise and capable of capturing multi-scale spectral features.  
The combination of BiLSTM and CNN allows the model to simultaneously exploit global and local information, thereby enhancing the recognition of B-type star spectral features. 

\begin{figure}[htbp]
    \centering
    
    \includegraphics[width=0.9\linewidth]{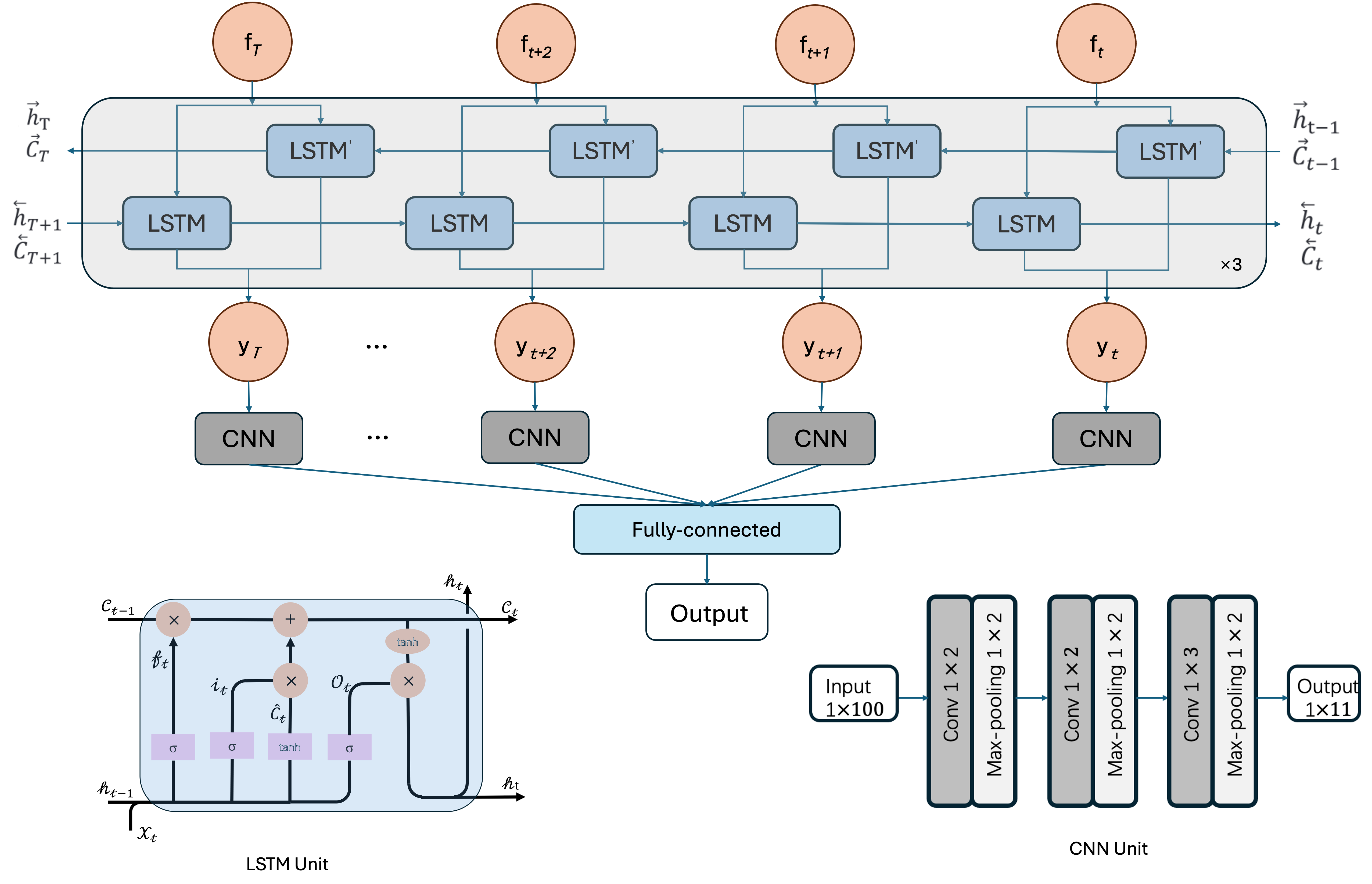}
    \caption{The structure of the BiLSTM-CNN model and the structure of the corresponding LSTM unit and CNN unit.}
    \label{model}
\end{figure}

Our model consists of two main components. First, three BiLSTM layers were applied to model the spectral sequences in both forward and backward directions, producing hidden state vectors that incorporate contextual information. The input of our model is a spectral segment of 1,800 flux values. To better adapt to the feature extraction of LSTM and convolutional layers, the input was reshaped into 36 arrays of size $1\times50$ (this configuration was chosen after testing several reshaping strategies).  
The output of the BiLSTM layers is a set of 36 arrays of size $1 \times 100$. These arrays were subsequently fed into a CNN layer to perform local convolution operations. The CNN layer consists of three convolutional layers combined with max-pooling layers, progressively extracting spectral features at different scales.  The overall architecture of our model is shown in Figure \ref{model}.

After feature extraction, the CNN output yields 36 arrays of size $1 \times 11$. These arrays were fused through fully connected layers, and the final classification was performed by a softmax layer \citep{NIPS1989Bridle}. For model training, we adopted cross-entropy \citep{cross} as the loss function and used the stochastic gradient descent \citep{hardt2016train} optimizer for parameter updates. To achieve stable model performance, multiple rounds of training and testing were conducted, and different hyperparameter configurations were compared. The optimal configuration was selected as the final model. 

We use the confusion matrix to evaluate our model. The confusion matrix consists of true positives (TP), false positives (FP), true negatives (TN), and false negatives (FN), where TP and TN represent correct predictions, and FP and FN represent incorrect predictions. The confusion matrix provides a clear view of how well the predicted labels match the true labels.
Additionally, from the confusion matrix, we can calculate accuracy, precision, recall, and F1 score\citep{forman2003} to further assess the model's performance for each category.

\begin{figure}[htbp]
\begin{center}
\includegraphics[width=0.5\textwidth]{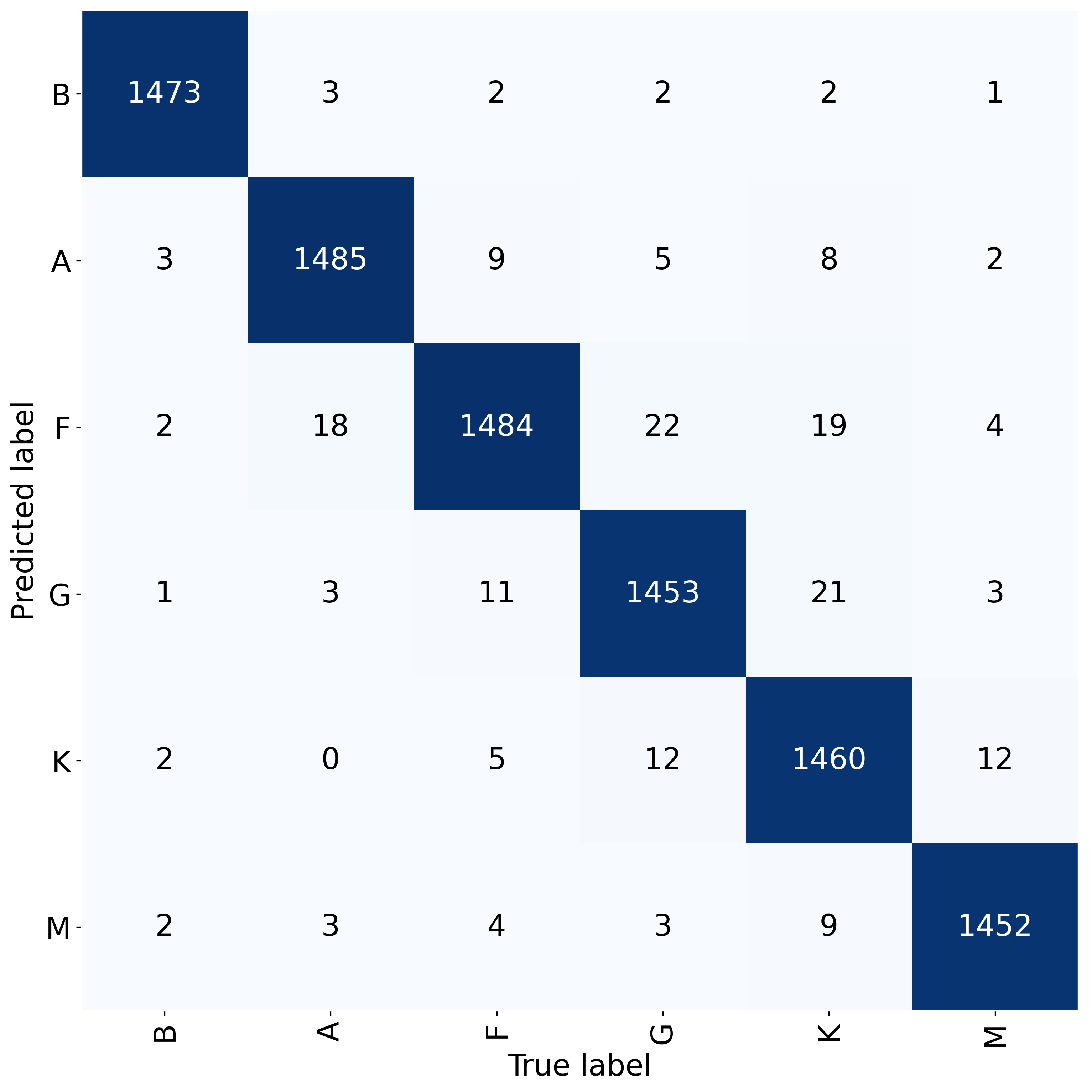}
\end{center}
\caption{Confusion matrix of the B-type star identification model.}
\label{Ha_cm}
\end{figure}

We set the learning rate to 0.001 and the batch size to 400, with a total of 2000 iterations. After the model training, we saved the model with the highest accuracy for subsequent selection of B-type star candidates. The model fitted after approximately 1900 iterations, achieving a maximum accuracy of 97.86\%, with the corresponding confusion matrix shown in Figure \ref{Ha_cm}. The precision rate, recall rate, and F1 scores for each category are shown in Table \ref{Ha}.

\begin{table}[htbp]
\centering
\caption{Accuracy, Precision, Recall and F1 score of the B-type star identification Model. }
\renewcommand\tabcolsep{20pt}
\begin{tabular}{ccccc}
\hline
Class & Precision & Recall & F1 & Accuracy\\
\hline 
B&99.33$\%$&99.33$\%$&99.33$\%$&\nodata \\
A&98.21$\%$&98.21$\%$&98.21$\%$&\nodata \\
F&95.80$\%$&97.95$\%$&96.87$\%$&\nodata \\
G&97.39$\%$&97.06$\%$&97.22$\%$&\nodata \\
K&97.92$\%$&96.12$\%$&97.01$\%$&\nodata \\
M&98.57$\%$&98.51$\%$&98.54$\%$&\nodata \\
All& \nodata &\nodata & \nodata&97.86$\%$ \\
\hline
\end{tabular}
\label{Ha}
\end{table}

\subsection{Acquisition of Be star Candidates}\label{3.2}


After model training and performance validation, we applied the model to the entire set of LAMOST DR11 low-resolution spectra. Under the S/N constraint (S/N $\geq$ 10), the model initially selected 55,667 B-type candidates. We first removed spectra that overlapped with the B-type star catalog published by \citet{Liu2019} (20,037 samples), leaving 34,630 spectra as new B-type candidates.

Subsequently, we combined the MKCLASS automatic spectral classification tool with visual inspection to verify all 34,630 spectra individually. Finally, 20186 B-type spectra were confirmed. Including the overlapping objects from Liu’s catalog, we identified a total of 40223 B-type stars. Table \ref{40223} presents the basic information of these B-type stars.

\begin{figure}[htbp]
\begin{center}
\includegraphics[width=0.85\textwidth]{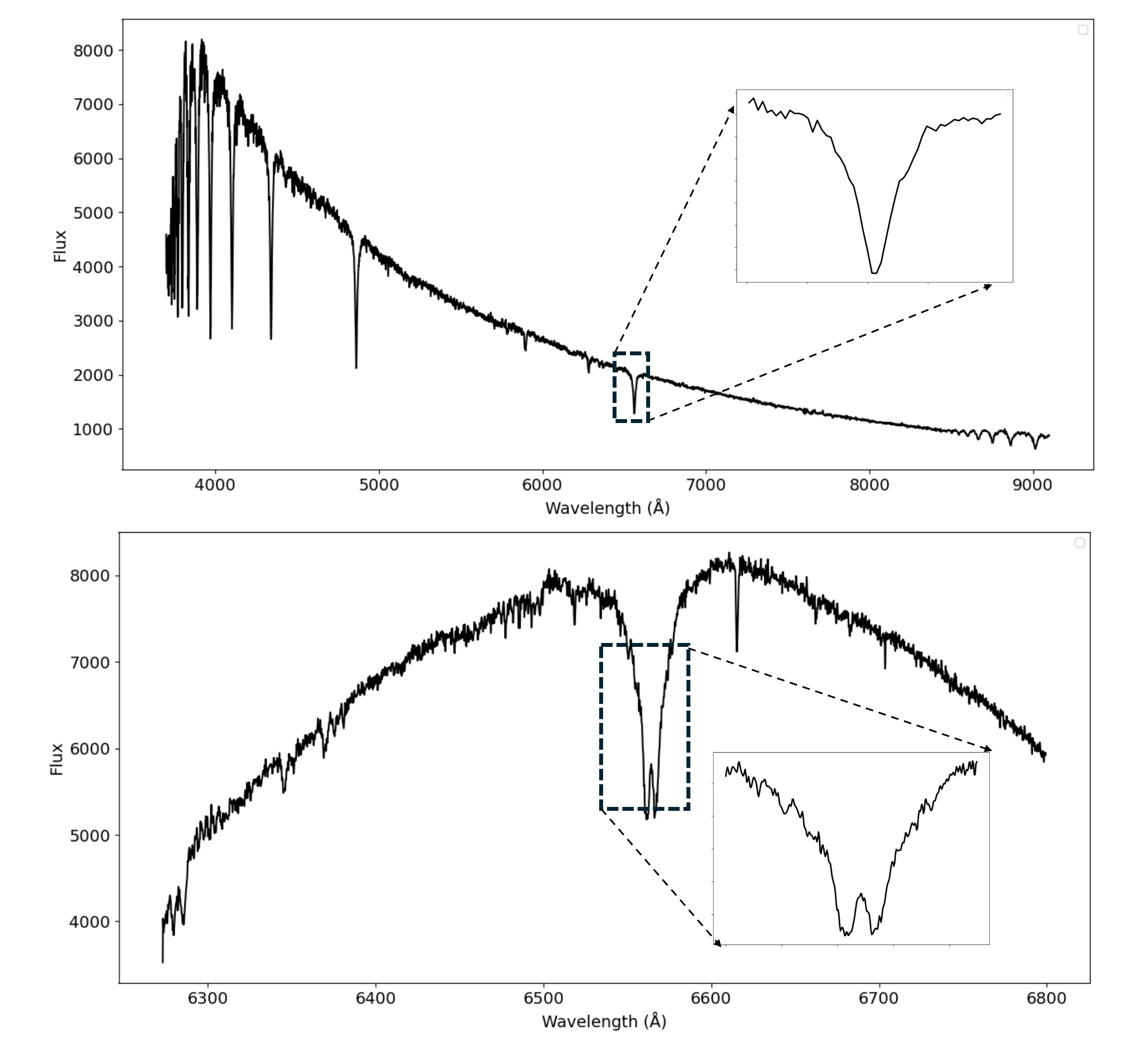}
\end{center}
\caption{Illustration of H$\alpha$ emission lines. The upper panel shows a low-resolution spectrum without a clear H$\alpha$ emission feature, while the lower panel displays the corresponding medium-resolution spectrum, in which the H$\alpha$ emission line is clearly visible.}
\label{compare}
\end{figure}

We then cross-matched the identified B-type stars with the H$\alpha$ emission-line star catalogs published by \citet{Tan2024}, \citet{Tan2025} , \citet{Hou2016} and \citet{Zhang2022}. We performed a visual inspection of the cross-matched results and removed contaminants such as white dwarfs and hot subdwarfs. Finally, a total of 8298 Be stars were identified.
It is worth noting that the catalog of \citet{Tan2025} is based on medium-resolution spectra. Due to the differences in spectral resolution, some spectra without prominent H$\alpha$ emission in the LAMOST low-resolution spectra show clear emission features in the medium-resolution data. Figure \ref{compare} presents an example, where the low-resolution spectrum shows no obvious H$\alpha$ emission line, while the corresponding medium-resolution spectrum reveals a distinct emission feature.

Finally, we compared these cross-matched Be star candidates with previously published Be star catalogs \citep{Vioque2018, Shridharan2021RAA, Lin2015, Hou2016, Chojnowski2015, Chen2016MNRAS, Zhang2022, Wang2022}. After removing known objects, we confirmed 4511 newly identified Be stars, among which 1656 have corresponding medium-resolution spectra. Table \ref{4512} presents the basic information of the newly identified Be stars.

\section{Classification and Validation of Be stars}\label{4}
We aimed to distinguish between CBe stars and HBe stars from the beginning of our analysis, as both classes exhibit H$\alpha$ emission features that can lead to classification ambiguity. Following \citet{Finkenzeller1984}, the HBe stars are known to show stronger infrared excesses compared to CBe stars. This distinction can be effectively characterized through infrared color indices. \citet{Hou2016AA} demonstrated that the traditional $(H-K, K-L)$ color--color diagram can be replaced by $(H-K, K-W1)$ when adopting the WISE $W1$ band, providing an efficient means to separate the two populations.  

Adopting this approach, we cross-matched our Be star candidates with the AllWISE Data Release \citep{Cutri2014yCat} and obtained corresponding $J$, $H$, $K$, and $W1$ photometric measurements. To account for the effects of interstellar extinction, we applied reddening corrections to all infrared magnitudes. The line-of-sight reddening values, $E(B-V)$, were derived from the three-dimensional dust map of \citet{Green2019ApJ} using the equatorial coordinates and distance estimates from \textit{Gaia} EDR3 \citep{Bailer2021AJ}. The adopted extinction ratios $A_\lambda/A_V$ for the $J$, $H$, and $K$ bands were taken from \citet{Cardelli1989ApJ}, assuming $R_V = A_V/E(B-V) = 3.1$ \citep{Fitzpatrick1999PASP}, with values of 0.282, 0.190, and 0.114, respectively. For the WISE $W1$ band, we used the coefficient $A_{W1}/A_V = 0.061$ from \citet{Yuan2013MNRAS}.  

Based on the color--color separation criteria proposed by \citet{Hou2016AA}, stars with $(H-K) > 0.4$ and $(K-W1) > 0.8$ were classified as HBe stars, whereas those with $(H-K) < 0.2$ and $(K-W1) < 0.5$ were classified as CBe stars. Applying these thresholds to our extinction-corrected photometry allowed us to clearly distinguish the two groups. This method provides a reliable and physically motivated framework for separating main-sequence Be stars from their pre-main-sequence counterparts in our sample.

Table \ref{4512} presents the final classification results. In total, we identified 3363 CBe stars and 35 HBe stars. In addition, 870 objects that do not meet the above criteria are labeled as ``Mixed'', and 18 possible HBe candidates that satisfy only one criterion [$(H-K) > 0.4$] are labeled as ``HBe?''. Furthermore, 225 sources lack sufficient infrared information for classification and are labeled as ``None''.

\section{Discussions}\label{5}

\subsection{Model Limitations}
Although the BiLSTM--CNN hybrid deep learning model adopted in this study achieved high accuracy during training and testing, certain misclassifications remain when applied to large-scale spectroscopic data. These misclassifications can be broadly summarized into the following categories.  

First, early A-type stars are sometimes misidentified as B-type stars. These stars share certain spectral similarities with B-type stars, particularly in the depth and width of Balmer absorption lines. In the absence of strong helium line features with sufficiently high S/N to serve as discriminants, the model tends to incorrectly classify some A-type stars as B-type stars. Although this type of error accounts for only a small fraction of the large data set, it inevitably reduces the purity of early-type star classification.  

\begin{figure}[htbp]
    \centering
    
    \subfigure[]{\includegraphics[width=0.9\textwidth]{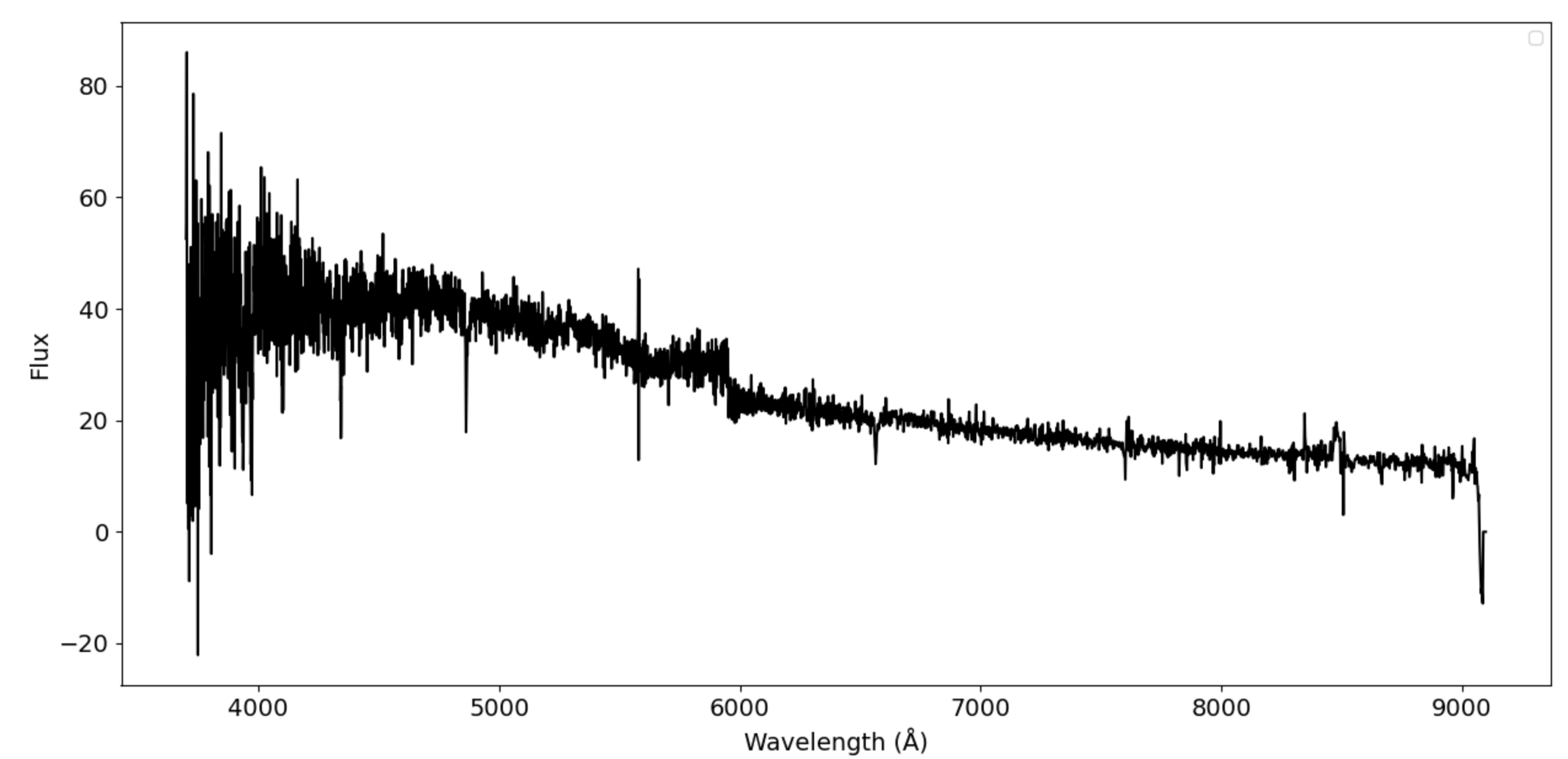}}
    
    \subfigure[]{\includegraphics[width=0.9\textwidth]{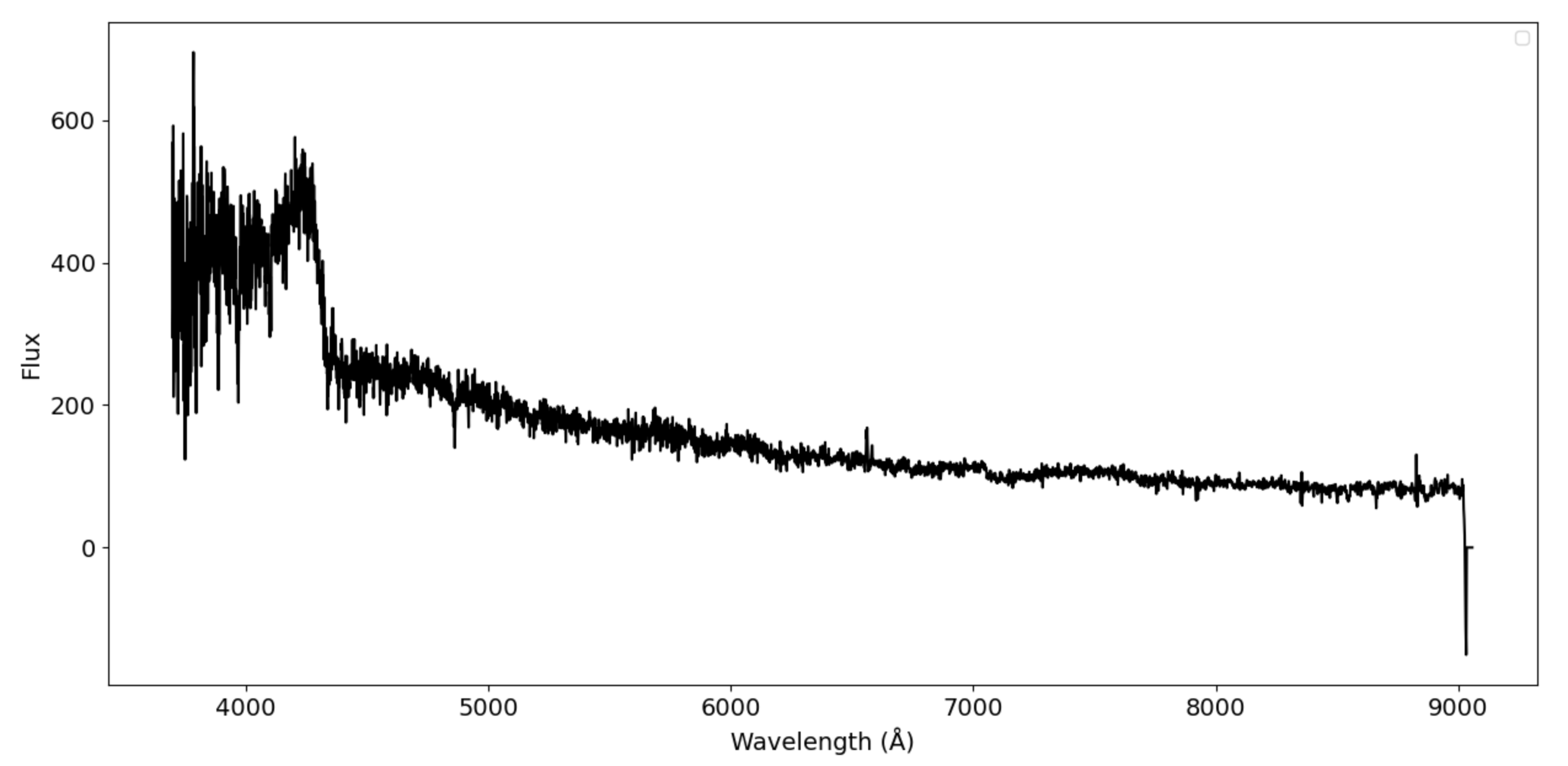}}

    \caption{Examples of samples misclassified by the B-type star identification model.}
    \label{misc}
\end{figure}

Second, spectra with relatively low S/N still pose challenges to the model. Despite applying a $g$-band S/N $\geq 10$ constraint during sample selection, some spectra still suffer from noise in the blue band. The noise obscures key spectral features, causing the local feature-extraction to become unreliable and resulting in misclassifications. Figure \ref{misc}(a) shows a sample with strong noise at the blue band of the spectrum, where the characteristic absorption lines of B-type stars are difficult to identify because they are buried in noise.

Finally, spectral anomalies induced by instrumental effects also contribute significantly to misclassifications. In some spectra, the continuum shows unreasonable depressions or bumps,and these features may mislead the model, causing it to produce incorrect predictions in regions of the feature space not well represented in the training set. Although we employed cubic spline fitting to remove the blackbody continuum and corrected for bad pixels during preprocessing, the model still struggles to handle extremely abnormal spectral morphologies with high accuracy. Figure \ref{misc}(b) presents a spectrum with an abnormal flux enhancement, exhibiting a pronounced bump around 4100\,\AA. Unlike cosmic-line hits or missing data points, such features are difficult to remove during preprocessing, and the identification and treatment of these spectra remain challenging.

\subsection{Samples analysis}\label{5.2}
\begin{figure}[htbp]
\begin{center}
\includegraphics[width=0.85\textwidth]{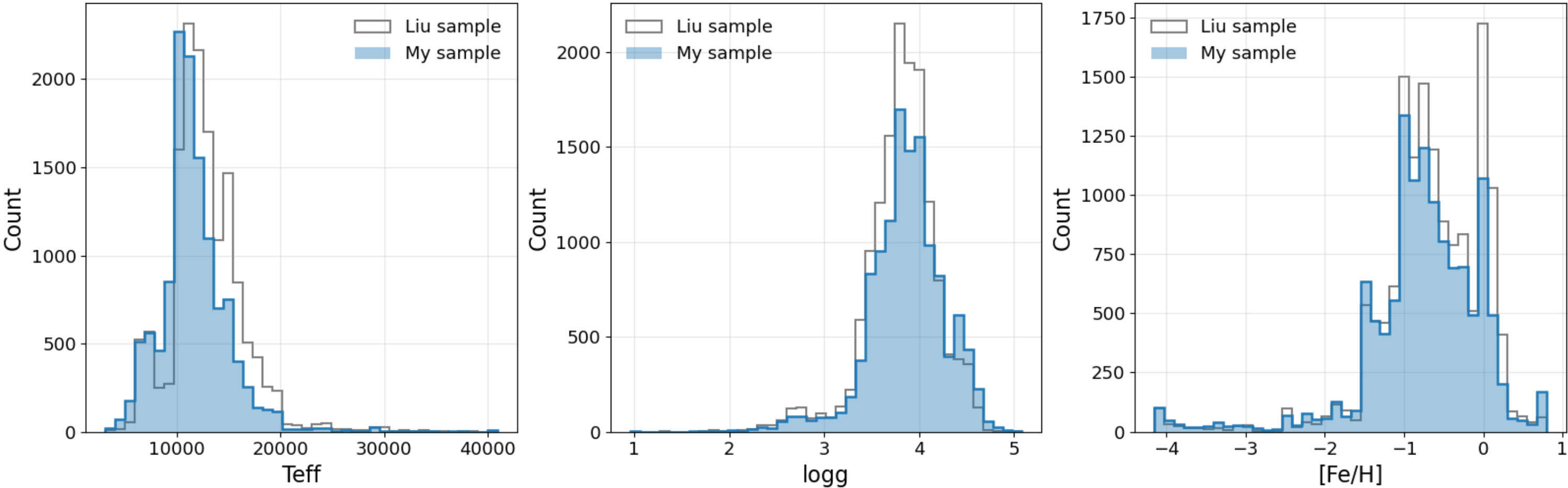}
\end{center}
\caption{Comparison of the stellar atmospheric parameter distributions between our B-type star sample (blue histograms) and the reference sample from \citet{Liu2019} (black outlines).}
\label{zhifang}
\end{figure}

We performed a detailed analysis of our identified B-type and Be star samples, starting with an examination of their stellar atmospheric parameters. Figure \ref{zhifang} compares the distributions of the stellar atmospheric parameters between our samples (blue histograms) and the reference samples from \citet{Liu2019} (black outlines). The effective temperature ($T_{\mathrm{eff}}$) distribution of our sample peaks around 10,000--12,000\,K, consistent with the typical range of B-type stars. The $\log g$ distribution exhibits a concentration near $\log g \approx 4.0$, indicating that most of our candidates are main-sequence stars. The [Fe/H] distribution displays a wide range from $-2.5$ to $+0.5$, with a peak around $-0.5$. As shown in the figure, the distributions of atmospheric parameters for our B-type star sample are consistent with those of \citet{Liu2019}. Overall, the similarity in the parameter distributions confirms the reliability of our classification model. 

\begin{figure}[htbp]
    \centering
    
    \subfigure[]{\includegraphics[width=0.45\textwidth]{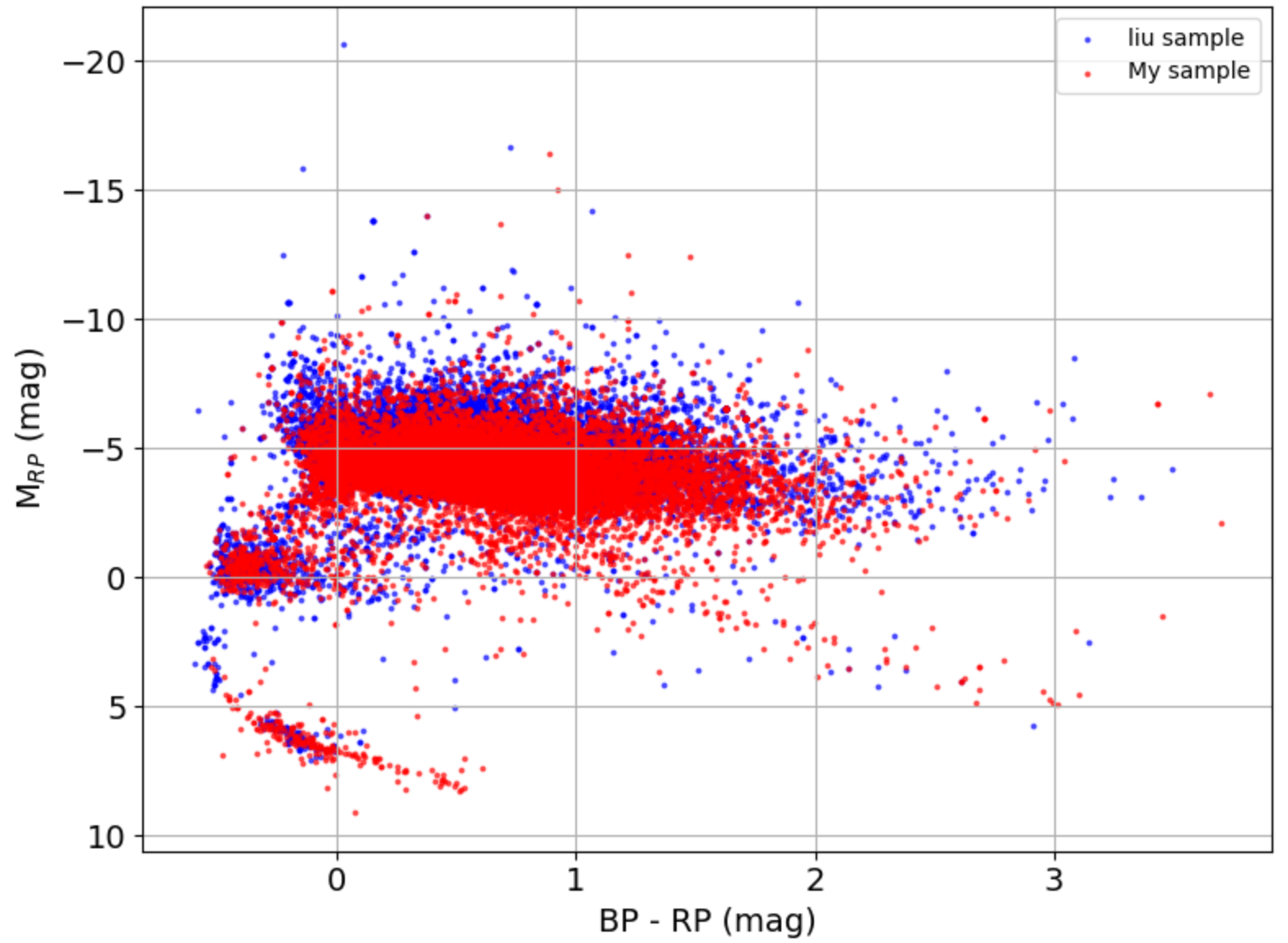}}
    \subfigure[]{\includegraphics[width=0.45\textwidth]{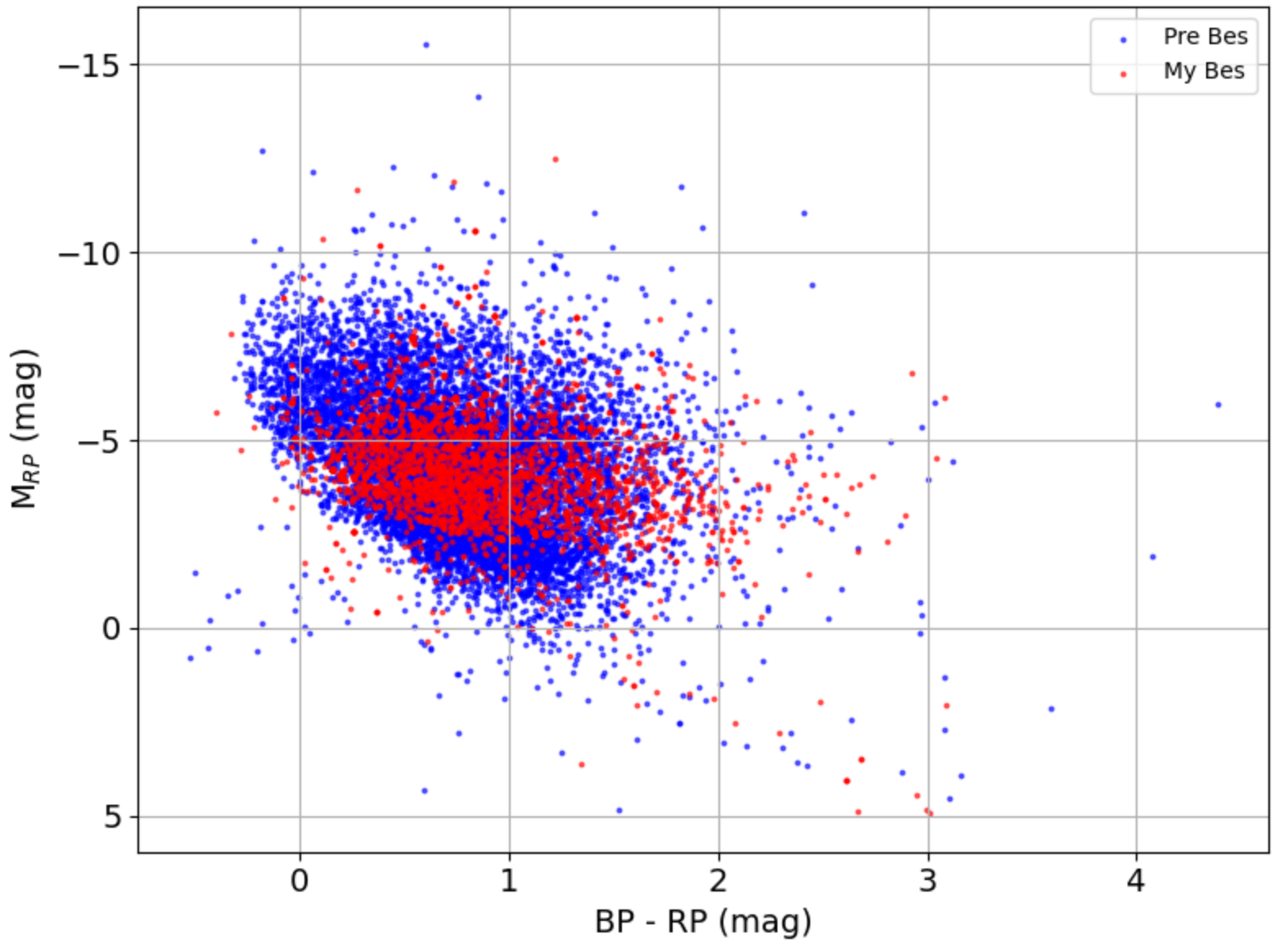}}

    \caption{Comparison of our B-type and Be stars with those from previous studies in the HR diagram. The red points represent our samples, while the blue points correspond to previously identified sources.}
    \label{HR}
\end{figure}

Furthermore, we projected our B-type stars and Be stars onto the Hertzsprung–Russell (HR) diagram. Figure \ref{HR}a presents the HR diagram of our B-type star candidates (red) compared with the reference sample from \citet{Liu2019} (blue). Most samples occupy the typical main-sequence locus of early-type stars, showing a tight distribution along the sequence from $(BP-RP)\approx0$ to 1.5 and $M_{RP}$ between $-10$ and $0$\,mag. The consistency in their distributions indicates that our classification model successfully selects genuine B-type stars without introducing significant systematic bias.
Figure \ref{HR}b displays the HR diagram for the Be star samples. Similar to the B-type distribution, our Be stars (red) largely overlap with the previously known Be stars \citep{Vioque2018, Shridharan2021RAA, Lin2015, Hou2016, Chojnowski2015, Chen2016MNRAS, Zhang2022, Wang2022} (blue points in Figure \ref{HR}b), concentrated around $(BP-RP)\approx0.5$ and $M_{RP}\approx-3$ to $-6$\,mag. The overall agreement between the two samples further confirms the reliability of our selection and classification strategy.

\begin{figure}[htbp]
\begin{center}
\includegraphics[width=0.85\textwidth]{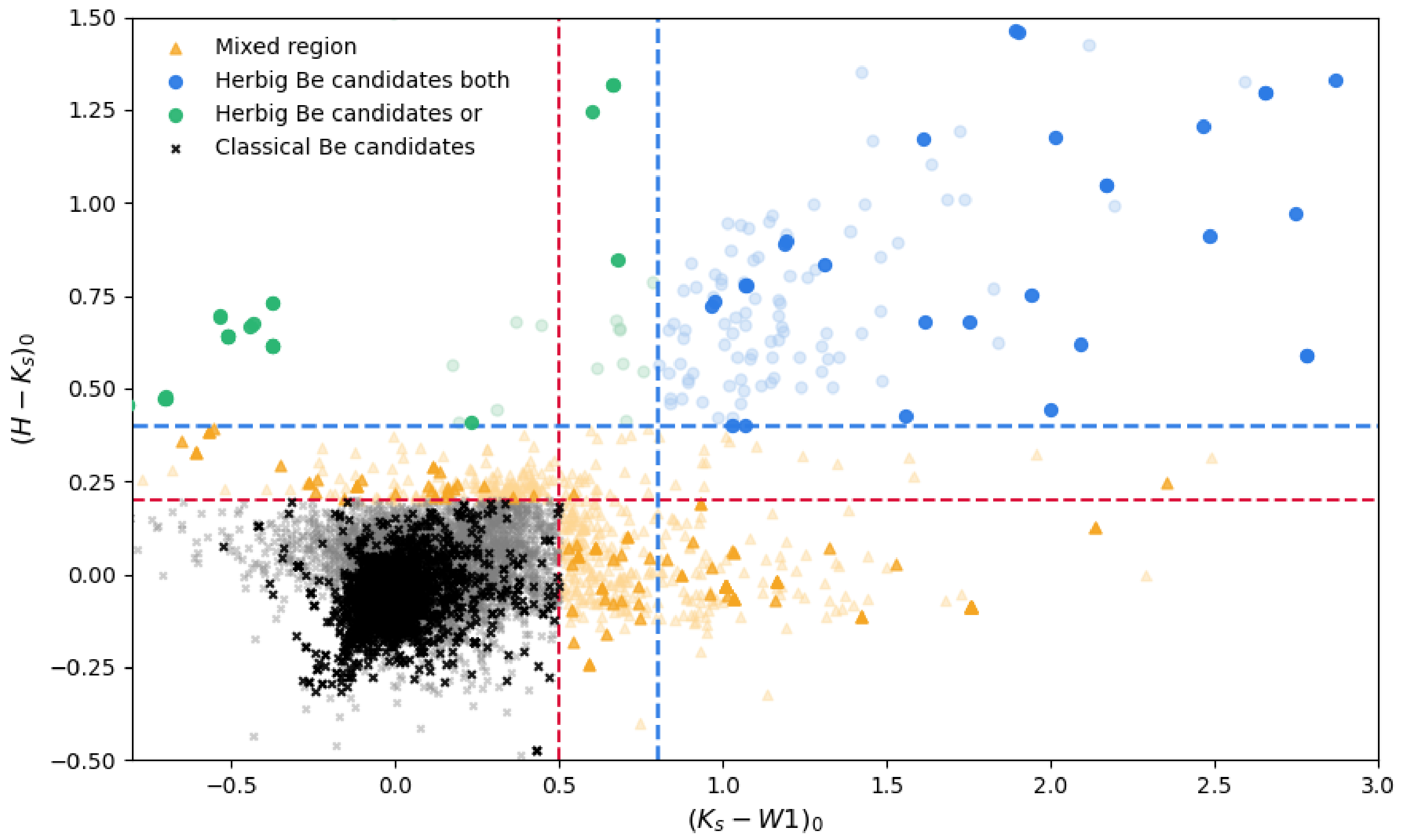}
\end{center}
\caption{Extinction-corrected $(H-K_s)_0$ versus $(K_s-W1)_0$ color--color diagram for the Be stars. All photometric measurements have been corrected for interstellar extinction. Black crosses represent CBe stars, blue symbols indicate HBe stars satisfying both selection criteria, yellow symbols denote ambiguous objects, and green symbols mark stars identified as HBe under only one of the criteria. The darker points correspond to our newly identified samples, while the lighter ones represent previously known sources.}
\label{JK}
\end{figure}

Figure \ref{JK} shows the color--color diagram of $(H-K_s)_0$ versus $(K_s-W1)_0$ used to separate CBe and HBe candidates. The dashed lines represent the empirical boundaries proposed by \citet{Hou2016AA}. Our newly identified sources are plotted with darker colors, and previously known samples are shown in lighter shades for comparison.  
Most of the CBe candidates (black crosses) are concentrated in the lower-left region of the diagram, exhibiting small infrared excesses typical of main-sequence Be stars. In contrast, the HBe candidates (blue circles) occupy the upper-right region, characterized by larger color indices indicating stronger circumstellar dust emission. A few objects fall into the intermediate zone (orange triangles), suggesting either uncertain classification or transitional properties. The close spatial overlap between the newly identified (dark symbols) and previously known (light symbols) sources demonstrates the consistency and reliability of our infrared-based classification method.

\section{Conclusions}\label{6}
In this study, we utilized the large-scale spectroscopic data from LAMOST DR11 to construct and apply a BiLSTM-CNN based deep learning model for a systematic search of Be stars. To ensure the stability and physical consistency of input features, we performed spectral preprocessing (selecting 1,800 blue-region pixels, removing continua, correcting cosmic rays and missing data). We then divided the data set into training (70\%) and testing (30\%) subsets and determined the optimal network architecture through experimentation. After training, the model was applied to all LAMOST DR11 spectra with $g$-band S/N $\geq 10$, yielding 55,667 B-type candidates. We removed overlaps with known samples and verified the remainder using MKCLASS \citep{Gray2014AJ} and manual inspection, confirming a total of 40,223 B-type stars (whih 20,186 newly discovered).

On this basis, we further cross-matched our B-type stars with the H$\alpha$ emission-line star catalogs published by \citet{Tan2024}, \citet{Tan2025} , \citet{Hou2016} and \citet{Zhang2022}. Through cross-matching and visual inspection, we identified a total of 8298 Be stars. After removing those previously reported in the literature, 4511 newly discovered Be stars were obtained. Based on their infrared color information, these new Be stars were further classified into CBe and HBe.

We compared the atmospheric parameters and positions in the HR diagram of our newly identified B-type and Be stars with those of previously published samples. The results show a high level of consistency between our new detections and the reference data. Furthermore, a comparison of the new Be stars with those reported in previous studies in the color--color diagram reveals an overall agreement in their distribution, confirming the reliability of our classification and selection methods.

Nevertheless, certain limitations remain. First, some early A-type stars were misclassified as B-type stars due to their spectral similarity in Balmer lines; without strong helium line features, the model struggles to discriminate between them. Second, although we imposed a threshold of S/N $\geq 10$, some spectra still suffer from noise, which obscures weak spectral features and leads to misclassification. Finally, instrumental effects occasionally caused abnormal continuum shapes, such as non-physical depressions or bumps, which could not be fully corrected even after spline fitting. To address these issues, we plan to explore more robust deep learning approaches, such as attention mechanisms, self-supervised learning, and multimodal feature fusion, to enhance stability in borderline and noisy cases. Incorporating high-resolution spectra and multi-band photometric data will further improve the validation and correction of identifications from low-resolution spectra, thereby enhancing both the accuracy and completeness of Be star searches. With future releases of LAMOST data and the continuous accumulation of survey data from Gaia, SDSS, and others, together with open-source tools and advancing deep learning techniques, we expect to further expand the Be star sample. 
The model training code, data processing scripts, and all associated catalogs are publicly available on Zenodo \citep{tan_2025_17454335}, providing convenient access for the broader research community.


\section{Acknowledgments}\label{7}

This work is supported by the China Natural Science Foundation (12433012, 12373097), the Basic and Applied Basic Research Foundation Project of Guangdong Province (2024A1515011503). This work is also supported by Guangzhou University Postgraduate Innovation Ability Cultivation Programme.


\begin{table}[htbp]
  \centering
    \caption{Basic information of the 40,223 B-type stars.}
    \label{40223}
    \begin{tabular*}{\textwidth}{@{\extracolsep{\fill}}ccccccc}
      \toprule
      OBSID & Designation & RA$^{\circ}$ & Dec$^{\circ}$ & snrg & Gaia DR3 ID  & label\\
      \midrule
215136&J030022.77+002836.2&45.0948822&0.4767484&701.51&30343944744320&Liu \\
303002&J051956.43+282001.6&79.9851322&28.3337951&44.16&3421599032233447168&New \\
305185&J051749.28+281436.0&79.4553598&28.2433444&23.29&3422321858050605696&Liu \\
307179&J052942.98+274723.7&82.4290922&27.7899235&69.97&3442292352167957376&Liu \\
309024&J052728.81+285945.7&81.870043&28.9960544&25.72&3445687987672044288&Liu \\
310223&J051656.49+282749.2&79.2353936&28.4636825&69.68&3422387038474051328&Liu \\
313092&J053116.73+285527.4&82.8197117&28.9242879&18.59&3445836082441741056&Liu \\
415158&J004139.06+420713.5&10.412762&42.120425&22.11&381370731998505472&Liu \\
503194&J051920.00+284442.3&79.8333589&28.7450921&26.09&3422376867991583104&New \\
507179&J052942.98+274723.7&82.4290922&27.7899235&40.6&3442292352167957376&Liu \\
509072&J052652.56+300317.0&81.71902&30.054749&10.01&3446193080123178496&New \\
510158&J051616.52+275854.5&79.0688616&27.9818201&13.77&3422228743159370752&New \\
513098&J053116.73+285527.4&82.8197117&28.9242879&28.77&3445836082441741056&Liu \\
514062&J051543.43+290019.0&78.930966&29.005304&13.97&3422610479853946752&New \\
614086&J220448.74+313401.7&331.203121&31.567165&104.86&1898657883100188544&Liu \\
902031&J062407.76+264155.1&96.03234&26.698665&74.75&3432227461384464640&New \\
902090&J062231.17+262932.4&95.629903&26.492341&66.06&3432198530484903168&Liu \\
902097&J062257.77+262622.7&95.740732&26.439665&44.66&3432194574818800640&Liu \\
902129&J062419.14+264137.8&96.079767&26.693836&75.74&3432217153462945408&Liu \\
902155&J062156.37+264903.4&95.48489&26.817633&36.0&3432288449919841920&New \\
        ...& ...& ...& ...& ...& ... & ... \cr
      \bottomrule
    \end{tabular*}
\begin{flushleft}
\textbf{Notes.} Columns 1$\sim$5: Basic information from the LAMOST catalog. Column 6: Gaia DR3 ID.  Column 7: Results from visual inspection and MKCLASS. ``Liu'' denotes the samples from \citet{Liu2019}, while ``New'' refers to the newly identified sources in this work.
\end{flushleft}

\end{table}

\begin{table}[htbp]
    \caption{Basic information of the 4,512 Be stars.}
    \label{4512}
    \begin{tabular*}{\textwidth}{@{\extracolsep{\fill}}cccccccc}
      \toprule
      OBSID(Low) & OBSID(Med) & Designation & RA$^{\circ}$ & Dec$^{\circ}$ & snrg &Gaia DR3 ID & label \\
      \midrule
30214211&...&J013339.94+303826.1&23.41642&30.64061&24.63&303378317783394304&Mixed \\
31314004&...&J093512.20+310959.3&143.8008625&31.166475&30.54&697707958245002880&CBe \\
31804071&...&J041737.02+300740.2&64.404255&30.127843&14.79&165712109004835712&HBe? \\
32804097&...&J052439.71+282108.3&81.165466&28.352315&12.4&3445571478096422656&Mixed \\
32804236&...&J052551.51+285259.1&81.464626&28.883102&12.29&3445694241144237952&Mixed \\
33916103&...&J080726.79+303501.8&121.86165&30.58384&15.13&877023839284105600&Mixed \\
34205028&...&J034311.77+450912.9&55.799048&45.153587&460.48&244562554600186240&Mixed \\ 
34802181&...&J064746.56+053602.7&101.94401&5.6007709&32.8&3132402151005490560&Mixed \\
34814016&653307076&J064546.78+075101.3&101.44494&7.8503764&84.91&3134002588958755328&CBe \\
34814048&856705180&J064532.13+064507.9&101.3839137&6.7521963&27.74&3132784574893203968&CBe \\
34815159&...&J065126.24+080106.2&102.85937&8.0183895&18.87&3133353984477409280&CBe \\
34909096&856713037&J065728.90+075920.1&104.37043&7.988917&88.96&3157092612318301696&Mixed \\
34910200&...&J064536.83+054050.7&101.40347&5.6807639&11.37&3132373078368884224&CBe \\
35711014&...&J034513.32+521437.7&56.303389&52.243824&10.03&...&CBe \\
36205051&...&J061555.16+263030.9&93.979874&26.508586&26.43&3432949359488565504&CBe \\
36416125&...&J061622.73+285115.9&94.094735&28.854438&26.86&...&None \\
36702196&...&J133338.07+584933.7&203.4086458&58.8260361&60.5&1662154581230639232&CBe \\
36801089&...&J044552.08+451917.7&71.467012&45.321602&16.84&253018451851177856&CBe \\
67806095&...&J054327.19+272056.6&85.863294&27.34908&13.44&3441548330690799616&CBe \\
67808154&...&J053746.58+270554.8&84.444117&27.098569&56.02&3441307709442891776&HBe \\

        ... & ... & ... & ... & ...  & ... & ... & ... \cr
      \bottomrule
    \end{tabular*}
      \begin{flushleft}
      \textbf{Notes.} Columns 1$\sim$6: Basic information from the LAMOST catalog. Column 7: Gaia DR3 ID. Columns 8: Classification results for the Be stars. ``CBe'' and ``HBe'' represent objects that fully satisfy the respective classification criteria, ``Mixed'' indicates sources that do not meet either criterion, ``HBe?'' denotes those fulfilling only one of the HBe criteria, and ``None'' refers to objects lacking sufficient infrared color information for reliable classification.

      \end{flushleft}
\end{table}

\bibliographystyle{aasjournal}

\begin{thebibliography}{}
\expandafter\ifx\csname natexlab\endcsname\relax\def\natexlab#1{#1}\fi
\providecommand{\url}[1]{\href{#1}{#1}}
\providecommand{\dodoi}[1]{doi:~\href{http://doi.org/#1}{\nolinkurl{#1}}}
\providecommand{\doeprint}[1]{\href{http://ascl.net/#1}{\nolinkurl{http://ascl.net/#1}}}
\providecommand{\doarXiv}[1]{\href{https://arxiv.org/abs/#1}{\nolinkurl{https://arxiv.org/abs/#1}}}

\bibitem[{{Aidelman} {et~al.}(2020){Aidelman}, {Escudero}, {Ronchetti},
  {Quiroga}, \& {Lanzarini}}]{Aidelman2020}
{Aidelman}, Y., {Escudero}, C., {Ronchetti}, F., {Quiroga}, F., \& {Lanzarini},
  L. 2020, Communications in Computer and Information Science book series,
  1291, 111, \dodoi{10.1007/978-3-030-61218-4_8}

\bibitem[{{Anusha} {et~al.}(2021){Anusha}, {Mathew}, {Shridharan}, {Arun},
  {Nidhi}, {Banerjee}, {Kartha}, {Paul}, \& {Bhattacharyya}}]{Anusha2021}
{Anusha}, R., {Mathew}, B., {Shridharan}, B., {et~al.} 2021, \mnras, 501, 5927,
  \dodoi{10.1093/mnras/staa3964}

\bibitem[{{Bailer-Jones} {et~al.}(2021){Bailer-Jones}, {Rybizki}, {Fouesneau},
  {Demleitner}, \& {Andrae}}]{Bailer2021AJ}
{Bailer-Jones}, C.~A.~L., {Rybizki}, J., {Fouesneau}, M., {Demleitner}, M., \&
  {Andrae}, R. 2021, \aj, 161, 147, \dodoi{10.3847/1538-3881/abd806}

\bibitem[{Bridle(1989)}]{NIPS1989Bridle}
Bridle, J.~S. 1989, in Proceedings of the 3rd International Conference on
  Neural Information Processing Systems, NIPS'89 (Cambridge, MA, USA: MIT
  Press), 211–217

\bibitem[{{Cardelli} {et~al.}(1989){Cardelli}, {Clayton}, \&
  {Mathis}}]{Cardelli1989ApJ}
{Cardelli}, J.~A., {Clayton}, G.~C., \& {Mathis}, J.~S. 1989, \apj, 345, 245,
  \dodoi{10.1086/167900}

\bibitem[{{Chen} {et~al.}(2016){Chen}, {Liu}, \& {Shan}}]{Chen2016MNRAS}
{Chen}, P.~S., {Liu}, J.~Y., \& {Shan}, H.~G. 2016, \mnras, 463, 1162,
  \dodoi{10.1093/mnras/stw1757}

\bibitem[{{Chojnowski} {et~al.}(2015){Chojnowski}, {Whelan}, {Wisniewski},
  {Majewski}, {Hall}, {Shetrone}, {Beaton}, {Burton}, {Damke}, {Eikenberry},
  {Hasselquist}, {Holtzman}, {M{\'e}sz{\'a}ros}, {Nidever}, {Schneider},
  {Wilson}, {Zasowski}, {Bizyaev}, {Brewington}, {Brinkmann}, {Ebelke},
  {Frinchaboy}, {Kinemuchi}, {Malanushenko}, {Malanushenko}, {Marchante},
  {Oravetz}, {Pan}, \& {Simmons}}]{Chojnowski2015}
{Chojnowski}, S.~D., {Whelan}, D.~G., {Wisniewski}, J.~P., {et~al.} 2015, \aj,
  149, 7, \dodoi{10.1088/0004-6256/149/1/7}

\bibitem[{Cui {et~al.}(2012)Cui, Zhao, Chu, Li, Li, Zhang, Su, Yao, Wang, Xing,
  {et~al.}}]{cui2012large}
Cui, X.-Q., Zhao, Y.-H., Chu, Y.-Q., {et~al.} 2012, RAA, 12, 1197

\bibitem[{{Cutri} {et~al.}(2021){Cutri}, {Wright}, {Conrow}, {Fowler},
  {Eisenhardt}, {Grillmair}, {Kirkpatrick}, {Masci}, {McCallon}, {Wheelock},
  {Fajardo-Acosta}, {Yan}, {Benford}, {Harbut}, {Jarrett}, {Lake}, {Leisawitz},
  {Ressler}, {Stanford}, {Tsai}, {Liu}, {Helou}, {Mainzer}, {Gettngs},
  {Gonzalez}, {Hoffman}, {Marsh}, {Padgett}, {Skrutskie}, {Beck}, {Papin}, \&
  {Wittman}}]{Cutri2014yCat}
{Cutri}, R.~M., {Wright}, E.~L., {Conrow}, T., {et~al.} 2021, {VizieR Online
  Data Catalog: AllWISE Data Release (Cutri+ 2013)}, VizieR On-line Data
  Catalog: II/328. Originally published in: IPAC/Caltech (2013)

\bibitem[{{Deng} {et~al.}(2012){Deng}, {Newberg}, {Liu}, {Carlin}, {Beers},
  {Chen}, {Chen}, {Christlieb}, {Grillmair}, {Guhathakurta}, {Han}, {Hou},
  {Lee}, {L{\'e}pine}, {Li}, {Liu}, {Pan}, {Sellwood}, {Wang}, {Wang}, {Yang},
  {Yanny}, {Zhang}, {Zhang}, {Zheng}, \& {Zhu}}]{deng2012lamost}
{Deng}, L.-C., {Newberg}, H.~J., {Liu}, C., {et~al.} 2012, Research in
  Astronomy and Astrophysics, 12, 735, \dodoi{10.1088/1674-4527/12/7/003}

\bibitem[{{Finkenzeller} \& {Mundt}(1984)}]{Finkenzeller1984}
{Finkenzeller}, U., \& {Mundt}, R. 1984, \aaps, 55, 109

\bibitem[{{Fitzpatrick}(1999)}]{Fitzpatrick1999PASP}
{Fitzpatrick}, E.~L. 1999, \pasp, 111, 63, \dodoi{10.1086/316293}

\bibitem[{Forman {et~al.}(2003)}]{forman2003}
Forman, G., {et~al.} 2003, J. Mach. Learn. Res., 3, 1289

\bibitem[{{Gkouvelis} {et~al.}(2016){Gkouvelis}, {Fabregat}, {Zorec},
  {Steeghs}, {Drew}, {Raddi}, {Wright}, \& {Drake}}]{Gkouvelis2016}
{Gkouvelis}, L., {Fabregat}, J., {Zorec}, J., {et~al.} 2016, \aap, 591, A140,
  \dodoi{10.1051/0004-6361/201527090}

\bibitem[{{Gray} \& {Corbally}(2014)}]{Gray2014AJ}
{Gray}, R.~O., \& {Corbally}, C.~J. 2014, \aj, 147, 80,
  \dodoi{10.1088/0004-6256/147/4/80}

\bibitem[{{Green} {et~al.}(2019){Green}, {Schlafly}, {Zucker}, {Speagle}, \&
  {Finkbeiner}}]{Green2019ApJ}
{Green}, G.~M., {Schlafly}, E., {Zucker}, C., {Speagle}, J.~S., \&
  {Finkbeiner}, D. 2019, \apj, 887, 93, \dodoi{10.3847/1538-4357/ab5362}

\bibitem[{Hardt {et~al.}(2016)Hardt, Recht, \& Singer}]{hardt2016train}
Hardt, M., Recht, B., \& Singer, Y. 2016, in Proceedings of the 33rd
  International Conference on International Conference on Machine Learning -
  Volume 48, ICML'16 (JMLR.org), 1225–1234

\bibitem[{{Herbig}(1960)}]{Herbig1960}
{Herbig}, G.~H. 1960, \apjs, 4, 337, \dodoi{10.1086/190050}

\bibitem[{{Hinton} \& {Salakhutdinov}(2006)}]{cross}
{Hinton}, G.~E., \& {Salakhutdinov}, R.~R. 2006, Science, 313, 504,
  \dodoi{10.1126/science.1127647}

\bibitem[{Hochreiter \& Schmidhuber(1997)}]{Hochreiter1997}
Hochreiter, S., \& Schmidhuber, J. 1997, Neural Computation, 9, 1735,
  \dodoi{10.1162/neco.1997.9.8.1735}

\bibitem[{Hou {et~al.}(2016)Hou, Luo, Hu, Li, Bai, \& Zhao}]{Hou2016}
Hou, W., Luo, A., Hu, J., {et~al.} 2016, Research in Astronomy and
  Astrophysics, 16, 138, \dodoi{10.1088/1674-4527/16/9/138}

\bibitem[{{Hou} {et~al.}(2016){Hou}, {Li}, \& {Zhang}}]{Hou2016AA}
{Hou}, Y.~J., {Li}, T., \& {Zhang}, J. 2016, \aap, 592, A138,
  \dodoi{10.1051/0004-6361/201628851}

\bibitem[{{Lin} {et~al.}(2015){Lin}, {Hou}, {Chen}, {Shao}, {Zhong}, \&
  {Yu}}]{Lin2015}
{Lin}, C.-C., {Hou}, J.-L., {Chen}, L., {et~al.} 2015, Research in Astronomy
  and Astrophysics, 15, 1325, \dodoi{10.1088/1674-4527/15/8/015}

\bibitem[{{Liu} {et~al.}(2019){Liu}, {Cui}, {Liu}, {Huang}, {Zhao}, \&
  {Zhang}}]{Liu2019}
{Liu}, Z., {Cui}, W., {Liu}, C., {et~al.} 2019, \apjs, 241, 32,
  \dodoi{10.3847/1538-4365/ab0a0d}

\bibitem[{{Mathew} {et~al.}(2008){Mathew}, {Subramaniam}, \&
  {Bhatt}}]{Mathew2008}
{Mathew}, B., {Subramaniam}, A., \& {Bhatt}, B.~C. 2008, \mnras, 388, 1879,
  \dodoi{10.1111/j.1365-2966.2008.13533.x}

\bibitem[{{McSwain} \& {Gies}(2005)}]{McSwain2005}
{McSwain}, M.~V., \& {Gies}, D.~R. 2005, \apjs, 161, 118,
  \dodoi{10.1086/432757}

\bibitem[{{Mohr-Smith} {et~al.}(2017){Mohr-Smith}, {Drew}, {Napiwotzki},
  {Sim{\'o}n-D{\'\i}az}, {Wright}, {Barentsen}, {Eisl{\"o}ffel}, {Farnhill},
  {Greimel}, {Mongui{\'o}}, {Kalari}, {Parker}, \& {Vink}}]{Mohr2017}
{Mohr-Smith}, M., {Drew}, J.~E., {Napiwotzki}, R., {et~al.} 2017, \mnras, 465,
  1807, \dodoi{10.1093/mnras/stw2751}

\bibitem[{{P{\'e}rez-Ortiz} {et~al.}(2017){P{\'e}rez-Ortiz},
  {Garc{\'\i}a-Varela}, {Quiroz}, {Sabogal}, \& {Hern{\'a}ndez}}]{Perez2017}
{P{\'e}rez-Ortiz}, M.~F., {Garc{\'\i}a-Varela}, A., {Quiroz}, A.~J., {Sabogal},
  B.~E., \& {Hern{\'a}ndez}, J. 2017, \aap, 605, A123,
  \dodoi{10.1051/0004-6361/201628937}

\bibitem[{{Porter} \& {Rivinius}(2003)}]{Porter2003}
{Porter}, J.~M., \& {Rivinius}, T. 2003, \pasp, 115, 1153,
  \dodoi{10.1086/378307}

\bibitem[{{Rivinius} {et~al.}(2013){Rivinius}, {Carciofi}, \&
  {Martayan}}]{Rivinius2013}
{Rivinius}, T., {Carciofi}, A.~C., \& {Martayan}, C. 2013, \aapr, 21, 69,
  \dodoi{10.1007/s00159-013-0069-0}

\bibitem[{{Sabogal} {et~al.}(2008){Sabogal}, {Mennickent}, {Pietrzy{\'n}ski},
  {Garc{\'\i}a}, {Gieren}, \& {Kolaczkowski}}]{Sabogal2008}
{Sabogal}, B.~E., {Mennickent}, R.~E., {Pietrzy{\'n}ski}, G., {et~al.} 2008,
  \aap, 478, 659, \dodoi{10.1051/0004-6361:20078418}

\bibitem[{Schuster \& Paliwal(1997)}]{Schuster}
Schuster, M., \& Paliwal, K. 1997, IEEE Transactions on Signal Processing, 45,
  2673, \dodoi{10.1109/78.650093}

\bibitem[{{Shridharan} {et~al.}(2021){Shridharan}, {Mathew}, {Nidhi}, {Anusha},
  {Arun}, {Kartha}, \& {Kumar}}]{Shridharan2021RAA}
{Shridharan}, B., {Mathew}, B., {Nidhi}, S., {et~al.} 2021, Research in
  Astronomy and Astrophysics, 21, 288, \dodoi{10.1088/1674-4527/21/11/288}

\bibitem[{{Tan} {et~al.}(2024){Tan}, {Liu}, {Wang}, {Mei}, {Wang}, {Deng}, \&
  {Liu}}]{Tan2024}
{Tan}, L., {Liu}, Z., {Wang}, X., {et~al.} 2024, \apjs, 273, 34,
  \dodoi{10.3847/1538-4365/ad5a08}

\bibitem[{Tan {et~al.}(2025)Tan, Mei, \& Wang}]{tan_2025_17454335}
Tan, L., Mei, Y., \& Wang, F. 2025, B-type star identification model and data
  sets.,  Zenodo, \dodoi{10.5281/zenodo.17454335}

\bibitem[{{Tan} {et~al.}(2025){Tan}, {Mei}, {Qian}, {Wang}, {Cui}, {Huang},
  {Wang}, {Deng}, {Liu}, \& {Chi}}]{Tan2025}
{Tan}, L., {Mei}, Y., {Qian}, J., {et~al.} 2025, \apjs, 280, 24,
  \dodoi{10.3847/1538-4365/adf4e6}

\bibitem[{{{\v{C}}otar} {et~al.}(2021){{\v{C}}otar}, {Zwitter}, {Traven},
  {Bland-Hawthorn}, {Buder}, {Hayden}, {Kos}, {Lewis}, {Martell}, {Nordlander},
  {Stello}, {Horner}, {Ting}, {{\v{Z}}erjal}, \& {Galah
  Collaboration}}]{Cotar2021}
{{\v{C}}otar}, K., {Zwitter}, T., {Traven}, G., {et~al.} 2021, \mnras, 500,
  4849, \dodoi{10.1093/mnras/staa2524}

\bibitem[{{Vieira} {et~al.}(2003){Vieira}, {Corradi}, {Alencar}, {Mendes},
  {Torres}, {Quast}, {Guimar{\~a}es}, \& {da Silva}}]{Vieira2003}
{Vieira}, S.~L.~A., {Corradi}, W.~J.~B., {Alencar}, S.~H.~P., {et~al.} 2003,
  \aj, 126, 2971, \dodoi{10.1086/379553}

\bibitem[{{Vioque} {et~al.}(2018){Vioque}, {Oudmaijer}, {Baines},
  {Mendigut{\'\i}a}, \& {P{\'e}rez-Mart{\'\i}nez}}]{Vioque2018}
{Vioque}, M., {Oudmaijer}, R.~D., {Baines}, D., {Mendigut{\'\i}a}, I., \&
  {P{\'e}rez-Mart{\'\i}nez}, R. 2018, \aap, 620, A128,
  \dodoi{10.1051/0004-6361/201832870}

\bibitem[{{Vioque} {et~al.}(2020){Vioque}, {Oudmaijer}, {Schreiner},
  {Mendigut{\'\i}a}, {Baines}, {Mowlavi}, \&
  {P{\'e}rez-Mart{\'\i}nez}}]{Vioque2020}
{Vioque}, M., {Oudmaijer}, R.~D., {Schreiner}, M., {et~al.} 2020, \aap, 638,
  A21, \dodoi{10.1051/0004-6361/202037731}

\bibitem[{{{\v{S}}koda} {et~al.}(2020){{\v{S}}koda}, {Podsztavek}, \&
  {Tvrd{\'\i}k}}]{Skoda2020}
{{\v{S}}koda}, P., {Podsztavek}, O., \& {Tvrd{\'\i}k}, P. 2020, \aap, 643,
  A122, \dodoi{10.1051/0004-6361/201936090}

\bibitem[{{Wang} {et~al.}(2022){Wang}, {Li}, {Wu}, {Gies}, {Liu}, {Liu}, {Guo},
  {Chen}, \& {Han}}]{Wang2022}
{Wang}, L., {Li}, J., {Wu}, Y., {et~al.} 2022, \apjs, 260, 35,
  \dodoi{10.3847/1538-4365/ac617a}

\bibitem[{{Waters} \& {Waelkens}(1998)}]{Waters1998}
{Waters}, L.~B.~F.~M., \& {Waelkens}, C. 1998, \araa, 36, 233,
  \dodoi{10.1146/annurev.astro.36.1.233}

\bibitem[{{Witham} {et~al.}(2008){Witham}, {Knigge}, {Drew}, {Greimel},
  {Steeghs}, {G{\"a}nsicke}, {Groot}, \& {Mampaso}}]{Witham2008}
{Witham}, A.~R., {Knigge}, C., {Drew}, J.~E., {et~al.} 2008, \mnras, 384, 1277,
  \dodoi{10.1111/j.1365-2966.2007.12774.x}

\bibitem[{{Yuan} {et~al.}(2013){Yuan}, {Liu}, \& {Xiang}}]{Yuan2013MNRAS}
{Yuan}, H.~B., {Liu}, X.~W., \& {Xiang}, M.~S. 2013, \mnras, 430, 2188,
  \dodoi{10.1093/mnras/stt039}

\bibitem[{{Zhang} {et~al.}(2022){Zhang}, {Hou}, {Luo}, {Li}, {Qin}, {Lu}, {Li},
  {Chen}, \& {Zhao}}]{Zhang2022}
{Zhang}, Y.-J., {Hou}, W., {Luo}, A.~L., {et~al.} 2022, \apjs, 259, 38,
  \dodoi{10.3847/1538-4365/ac4964}

\bibitem[{{Zhao} {et~al.}(2012){Zhao}, {Zhao}, {Chu}, {Jing}, \&
  {Deng}}]{zhao2012lamost}
{Zhao}, G., {Zhao}, Y.-H., {Chu}, Y.-Q., {Jing}, Y.-P., \& {Deng}, L.-C. 2012,
  RAA, 12, 723, \dodoi{10.1088/1674-4527/12/7/002}

\bibitem[{{Zorec} \& {Briot}(1997)}]{Zorec1997}
{Zorec}, J., \& {Briot}, D. 1997, \aap, 318, 443

\end{thebibliography}

\end{sloppypar}
\end{CJK*}
\end{document}